\documentclass[reprint,prb]{revtex4-1}
\usepackage{graphicx,epsf,epsfig}
\usepackage{subfigure}
\usepackage{dcolumn}
\usepackage{bm}
\usepackage{textcomp}
\usepackage{amsmath}
\usepackage[normalem]{ulem} 

\begin{document}

\title{Modified Dirac Hamiltonian for Efficient Quantum Mechanical Simulations
of Micron Sized Devices}
\author{K. M. Masum Habib}
\email{masum.habib@virginia.edu}
\affiliation{Department of Electrical and Computer Engineering, 
University of Virginia, Charlottesville, VA 22904.}
\author{Redwan N. Sajjad}
\affiliation{Department of Electrical Engineering and Computer Science, Massachusetts Institute of Technology, Cambridge, MA-02139}
\author{Avik W. Ghosh}
\affiliation{Department of Electrical and Computer Engineering, 
University of Virginia, Charlottesville, VA 22904.
}%

\begin{abstract} 


    Representing massless Dirac fermions on a spatial lattice poses
    a potential challenge known as the Fermion Doubling problem. Addition of 
    a quadratic term to the Dirac Hamiltonian circumvents this problem. 
    We show that the modified Hamiltonian with the additional term results 
    in a very small Hamiltonian matrix when discretized on a real 
    space square lattice. The resulting Hamiltonian matrix is considerably more 
    efficient for numerical simulations without sacrificing on accuracy and is several 
    orders of magnitude faster than the atomistic tight 
    binding model. Using this Hamiltonian and the Non-Equilibrium Green's 
    Function (NEGF) formalism, we show several transport phenomena in graphene, 
    such as magnetic focusing, chiral tunneling in the ballistic limit 
    and conductivity in the diffusive limit in micron sized graphene 
    devices. The modified Hamiltonian can be used for any system with massless Dirac 
    fermions such as Topological Insulators, opening up a simulation domain that is not readily accessible otherwise. 
\end{abstract}

\maketitle

Two-dimensional materials have attracted considerable recent attention for their 
superior electronic characteristics and potential for electronic, 
opto-electronic and spintronic applications \cite{wang2012,mellnik2014}. 
Among them, graphene, topological insulators and transition metal 
dichalcogenides stand out in particular. Experimental progress on 
such materials are advancing rapidly \cite{butler2013} and modeling carrier 
transport to capture the new physics of these materials has become quite 
crucial. NEGF based numerical calculation is widely used nowadays to accurately 
model nano-scale materials \cite{datta_97}. The formalism is extremely 
powerful in solving any quantum transport problem accurately including
sophisticated contact-channel effects and various forms of scatterings within a self-energy 
correction to the Hamiltonian. Despite these considerable strengths, NEGF has so far been 
considered mainly as a ballistic quantum transport simulation platform, since it  
becomes computationally prohibitive to model diffusive systems. This is especially true for experimentally relevant device dimensions which are often in the hundreds of nano-meters to $\mu$m regime.  

\begin{figure}
\includegraphics[width=3.25in]{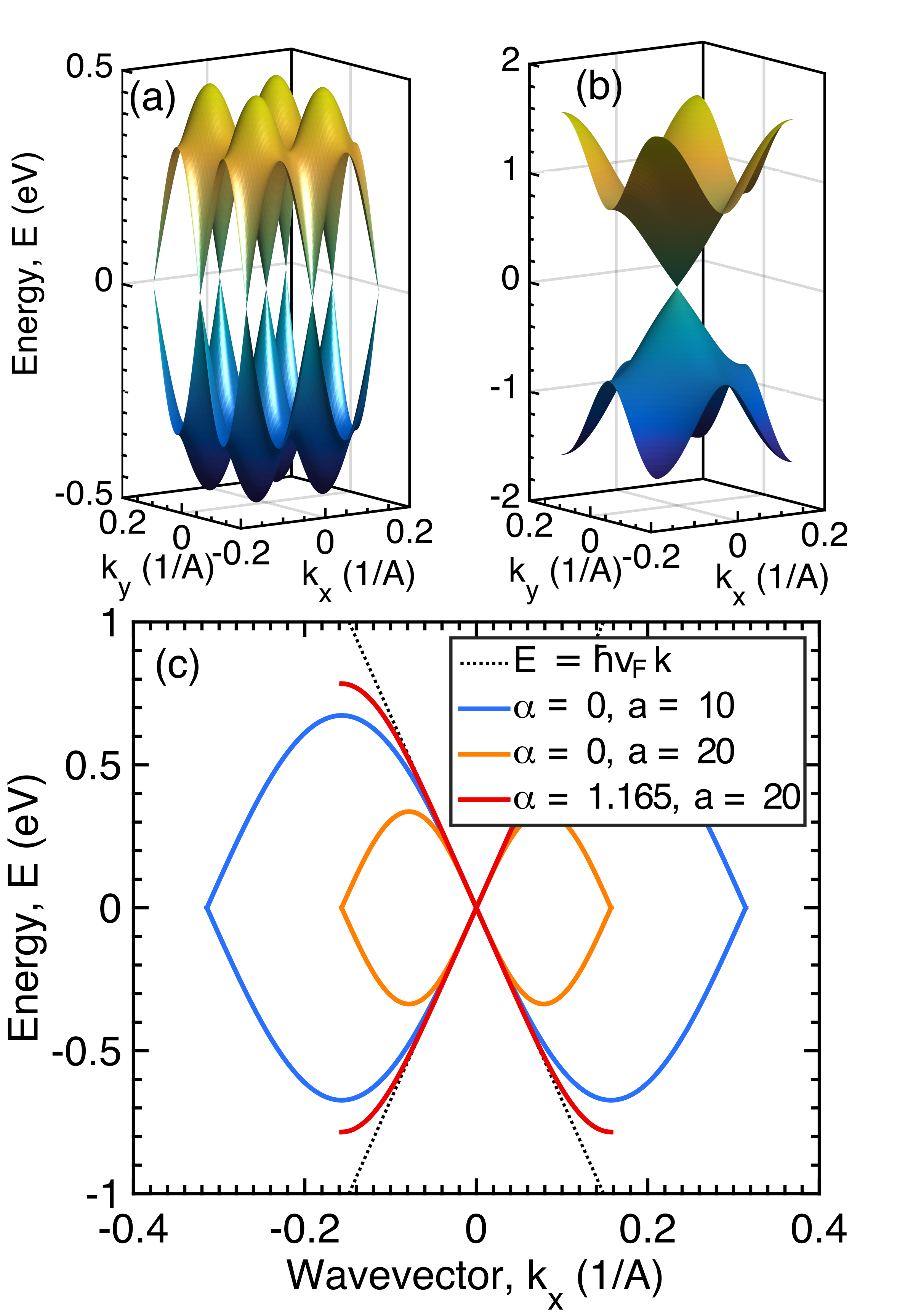}
\caption{The first Brillouin zone of graphene band structure calculated 
	using the discretized ${k.p}$ Hamiltonian in Eq. (2) for 
	(a) $\alpha = 0$ and (b) $\alpha = 1.165$. 
	(c) Comparison of band structures along $k_y=0$ for different $a$ and 
    $\alpha$. The lattice parameter $a$ is in \AA. With $\alpha = 1.165$,
     even a 2nm grid spacing results in a linear band structure for 
     $|E_\mathrm{max}| \sim 0.6$eV.
}\label{fig:band_kp}
\end{figure}

In materials such as graphene and topological insulators etc., electrons 
behave as massless Dirac fermions described by the Dirac Hamiltonian.
Quantum transport simulations using the NEGF formalism require a real 
space representation of the Hamiltonian matrix to fully describe the channel material. While the tight-binding representation is valid for most practical purposes, it is computationally expensive. 
To expedite the calculation, a discretized version of the Dirac Hamiltonian can be used. It has however been shown that representing the Dirac Fermions on a spatial lattice poses a problem commonly known as the Fermion Doubling \cite{Susskind_77,Stacy_PRD82} problem, where the numerical discretization of the Hamiltonian creates additional branches within the Brillouin zone (Fig. \ref{fig:band_kp}).
One way to solve this problem is to add a quadratic term with the Hamiltonian
\cite{Susskind_77, Hong_12}. 

\begin{figure}
\includegraphics[width=3in]{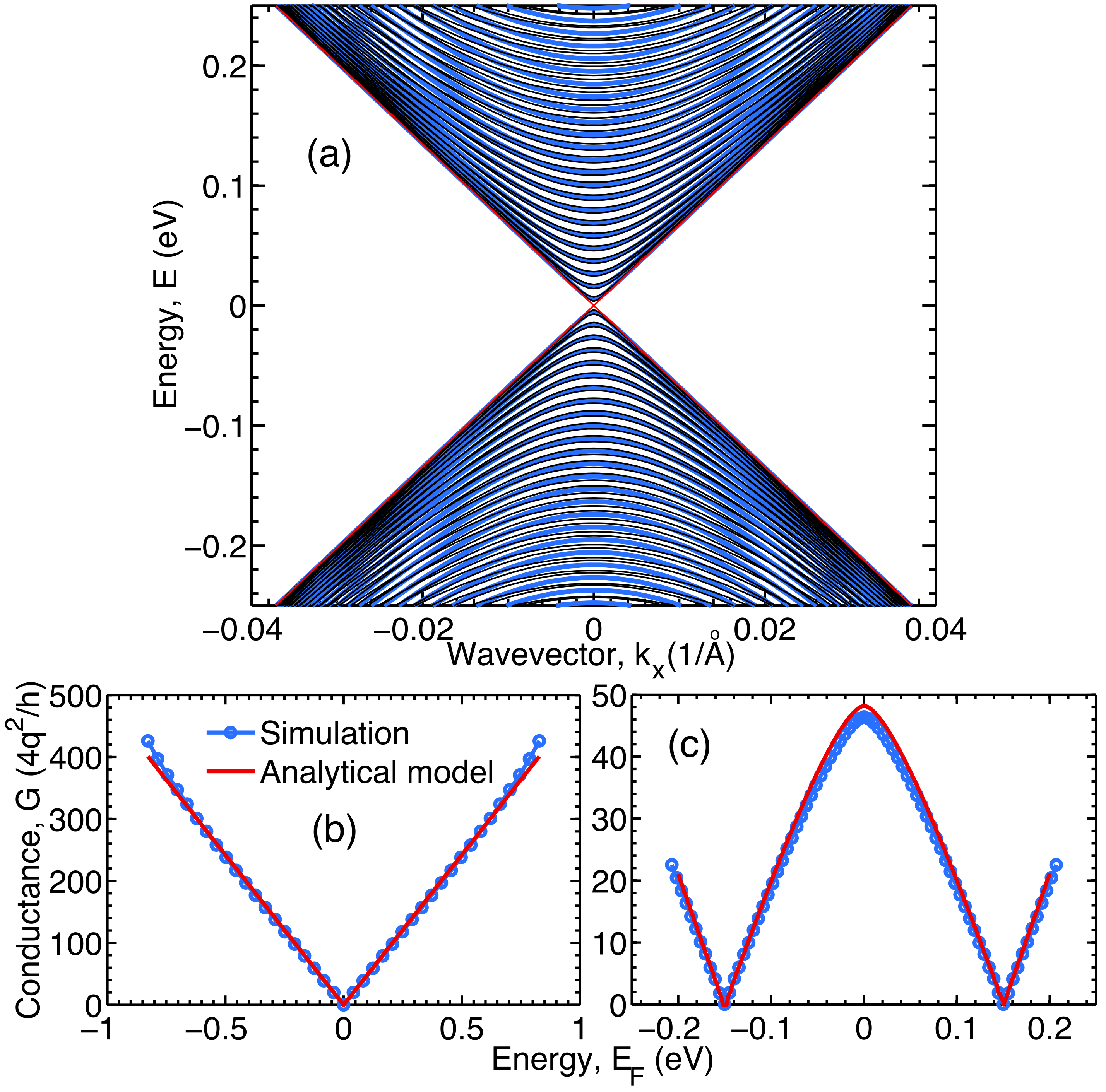}
\caption{(a) Band structure of 200nm armchair graphene nanoribbon calculated using the tight binding 
	model (in black) and the \emph{k.p} model (in blue) showing good agreement.
	The yellow line is the linear approximation of the graphene band structure. (b) Conductance of a $1\mu$m wide graphene sheet from NEGF simulation with the modified Hamiltonian along with that from linear $E-K$, (c) Conductance of a $1\mu$m wide graphene $pn$ junction in excellent agreement with exact analytical solutions, showing the model's ability to capture angle dependent chiral transmission.
}\label{fig:band_kp_ribbon}
\end{figure}

\begin{figure}
\centering
\includegraphics[width=3.25in]{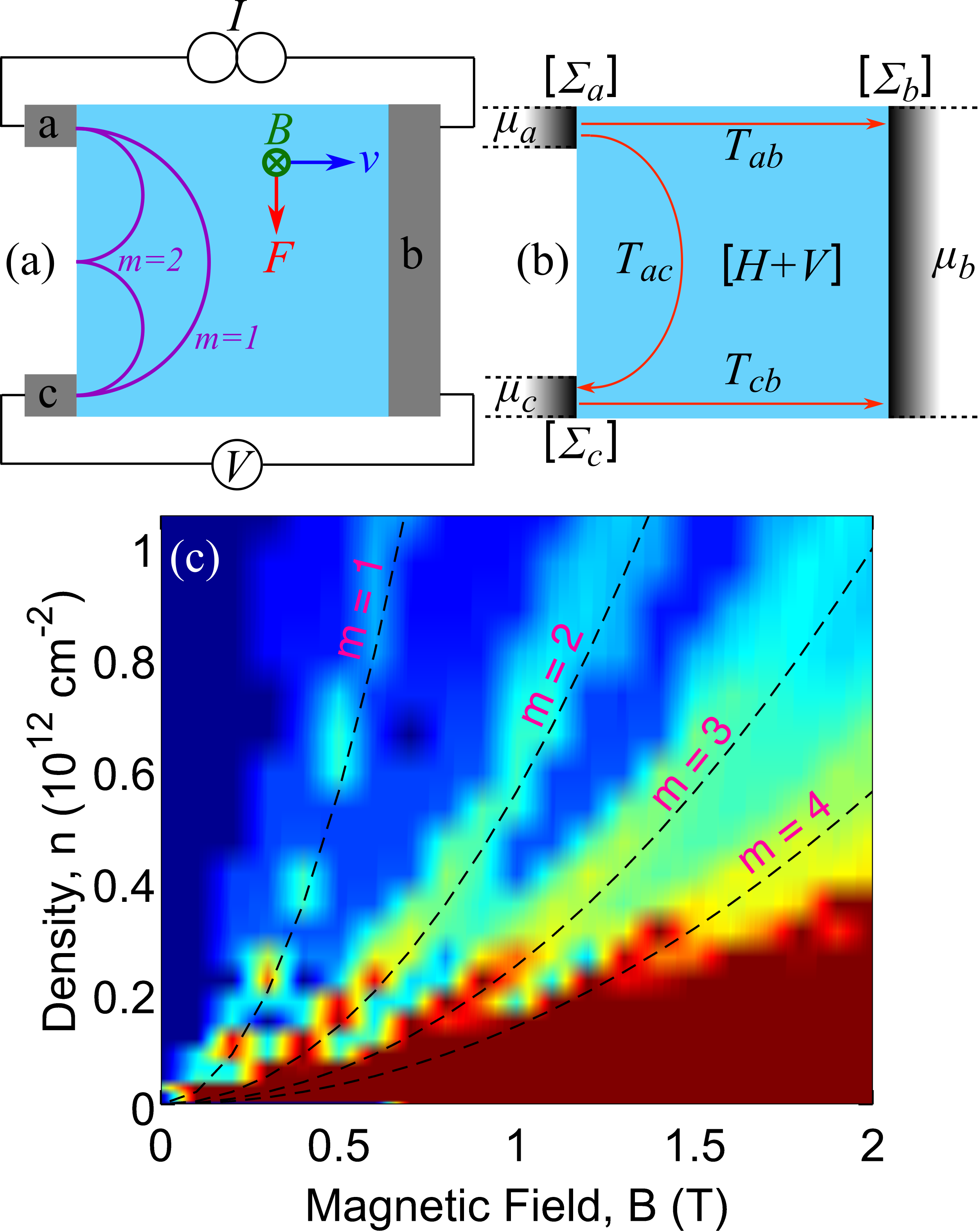}\quad
\caption{Large scale ballistic simulation to capture magnetic focusing 
    in graphene. (a) Device geometry and biasing scheme. Electrons
    are injected from contact $a$ and collected at contact $b$. The vertical 
    magnetic field $B$ forces the electrons to follow a circular path. For 
    some specific magnetic fields, electrons are focused to contact $c$, 
    resulting in a large voltage registered in the voltmeter. (b) The various matrix and transmission components that enter the NEGF simulation of the device. (c) The resistance, $R=V/I$ as a function of carrier density
    and magnetic field showing resonances. The dashed lines were calculated
    using Eq. (\ref{eq:B}). The channel dimensions are $\rm 400nm\times400nm$.}
\label{fig:focusing}
\end{figure}

\begin{figure}
\centering
\includegraphics[width=3.25in]{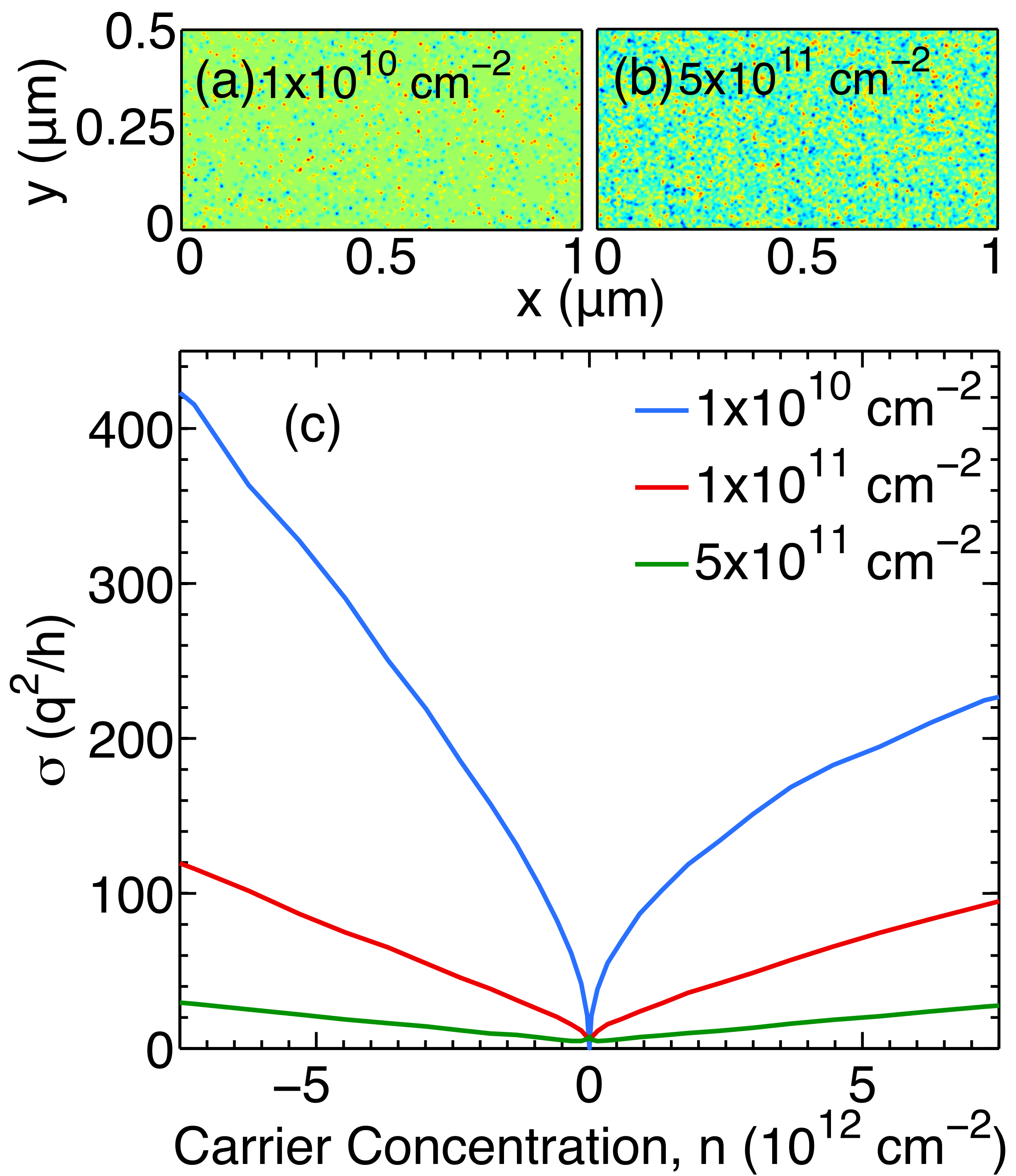}\quad
\caption{Modeling the transition from ballistic to diffusive transport in a micron long graphene, (a-b) sample potential landscape 
for various impurity concentrations, (c) total conductivity as a function of channel carrier 
density from ballistic ($\sigma \propto n$) to diffusive ($\sigma \propto \sqrt{n}$).}
\label{diff}
\end{figure}

In this letter, we show that the additional term not only solves the 
Fermion Doubling problem but also has a profound implication for numerical 
simulations. We show that the modified Dirac Hamiltonian results in 
a spatial Hamiltonian matrix which is several orders of magnitude smaller 
than the atomistic tight binding Hamiltonian \cite{wallace1947} while preserving the same level of accuracy over the relevant energy range. As a result, bandstructure and transport simulations using the modified 
Hamiltonian are several orders of magnitude faster than the tight binding,  
allowing us to carry out several large scale simulations such as angle dependent transmission in 1$\mu$m wide graphene (Fig. \ref{fig:band_kp_ribbon}), magnetic electron focusing in a $0.4\mu$m$\times0.4\mu$m multi-electrode graphene (Fig. \ref{fig:focusing}) and diffusive transport in a 1$\mu$m $\times$ 0.5$\mu$m graphene device (Fig. \ref{diff}) . Our results show very good agreement with recently
reported experimental results \cite{Herrero_focusing_NP13}. Other methods such as tight-binding make it almost impossible to simulate devices at such dimension. 

The modified effective ${k.p}$ Hamiltonian for graphene at low energy is,
\begin{eqnarray}
    H(k) = \hbar v_\mathrm{F} \left[k_x\sigma_x + k_y\sigma_y + 
    \beta(k_x^2 + k_y^2)\sigma_z\right] 
\label{eq:H_k}
\end{eqnarray}
where, $v_\mathrm{F}$ is the Fermi velocity, $\vec{k} = k_x\hat{x} + k_y\hat{y}$ 
is the wave vector, $\sigma$'s are Pauli matrices representing the 
pseudospins, and $\hbar$ is the reduced Planck's constant. The last term, 
$\beta(k_x^2 + k_y^2)\sigma_z$ serves two purposes: (i) it circumvents 
the well known Fermion doubling problem\cite{Stacy_PRD82} and (ii)
it allows us to generate a computationally efficient Hamiltonian
using a course grid with reasonable accuracy as shown below.
The k-space Hamiltonian in Eq. \ref{eq:H_k} is transformed to a real-space
Hamiltonian by replacing $k_x$ with differential operator 
$-i{\partial}/{\partial x}$, $k_x^2$ with $-{\partial^2}/{\partial x^2}$
and so on.
The differential operators are then discretized in a square lattice 
using the finite difference method to obtain, 
\begin{eqnarray}
\begin{split}
H = \sum_{i} c_i^\dagger \epsilon c_i 
+ \sum_{i} \left(c_{i,i}^\dagger t_x c_{i,i+1} + {\rm H. C.}\right)\\
+ \sum_{j} \left(c_{j,j}^\dagger t_y c_{j,j+1} + {\rm H. C.}\right)
\label{eq:H}
\end{split}
\end{eqnarray}
where $\epsilon = -4\hbar v_\mathrm{F}{\alpha}\sigma_z/{a}$, 
$t_x = \hbar v_\mathrm{F}\left[{i}\sigma_y/{2a} + {\alpha}\sigma_z/{a}\right]$,
$t_y = \hbar v_\mathrm{F}\left[-{i}\sigma_x/{2a} + {\alpha}\sigma_z/{a}\right]$,
$a$ is the grid spacing and $\alpha \equiv \beta/a$. 

When $\alpha = 0$, Eqs. (\ref{eq:H_k}) and (\ref{eq:H}) reduce to the
unmodified \emph{k.p} Hamiltonian. The eigen-energy calculated 
from Eq. \ref{eq:H} becomes a sine function of wavevector $\vec{k}$, and 
three extra Dirac cones appear inside the first Brillouin zone as shown 
in Fig. \ref{fig:band_kp}(a). 
This is known as the Fermion doubling problem\cite{Stacy_PRD82}. 
The last term in Eq. \ref{eq:H_k} gets rid of this problem by opening 
bandgaps for each of these Dirac cones. The resulting band structure 
contains only one Dirac cone as shown in Fig. \ref{fig:band_kp}(b).

The effect of the extra term in Eq. (\ref{eq:H_k}) is more clearly
illustrated in Fig. \ref{fig:band_kp}(c). When $\alpha=0$ and $a=20$A$^o$, 
the band structure along $k_y = 0$ is a sine function and an extra Dirac cone 
appears at the zone boundary. This Dirac cone is removed when $\alpha=1.165$
and the band structure closely follows the ideal band structure of graphene. 
The band structure calculated using $\alpha=1.165$ 
with $a = 20$\AA~is accurate over a larger window than that is calculated 
using $\alpha=0$ and $a=10$\AA. Thus, the $\alpha$ parameter not only 
circumvents the Fermion Doubling problem but also enables us to use coarser 
lattice grid without losing the accuracy. 
{With $a = 20$\AA~ and $\alpha=1.165$, the Hamiltonian size given by Eq. (\ref{eq:H})
for a 1$\mu {\rm m}\times1\mu {\rm m}$ graphene sheet is 
$5{\rm E}5\times5{\rm E}5$
compared to $\sim38{\rm E}6\times38{\rm E}6$ in atomistic tight binding model.}
The band structure calculated using this discretized Hamiltonian closely 
resembles the ideal linear band structure within $|E| \sim 0.6$eV.
To obtain the same level of accuracy, the recently proposed scaled graphene 
model\cite{Richter_PRL15} requires a Hamiltonian of size $\sim9{\rm E}6\times9{\rm E}6$.

The accuracy of the proposed Hamiltonian is demonstrated by the band structure 
of a 200nm graphene ribbon shown in Fig \ref{fig:band_kp_ribbon}. 
The bands calculated using the one $p_z$ orbital tight-binding model
(in black) and the modified Dirac Hamiltonian (in blue) are in good 
agreement in the low energy limit. For the tight binding model, the calculation
takes about 1 hour and 30 minutes. In comparison, the modified Hamiltonian
model takes about 2 seconds using the same number of cores. {This shows that our proposed model is three 
orders of magnitude faster with excellent accuracy for band structure
calculation.}
With the modified Dirac Hamiltonian, we employ the standard Recursive Green's 
Function (RGF) algorithm \cite{alam2005} and 
employ it for large scale simulations, in
both ballistic and diffusive regimes.

Fig. \ref{fig:band_kp_ribbon}b shows the conductance calculated using this model (with $\alpha = 1.16$ and $a = 16 A^0$) for a 1$\mu$m wide graphene sheet and compared with the conductance from the linear graphene $E-K$ relationship. We see a good agreement until $E_\mathrm{F} = 0.6$eV. In Fig. \ref{fig:band_kp_ribbon}c, we show the conductance of a electrostatically doped split-gated graphene $pn$ junction, showing excellent agreement with the
exact analytical solution \cite{sajjad_13,sajjad_12}. That means the angle dependent transmission, an important property of graphene originating from its chiral nature \cite{katsnelson_06}, is undistorted in the modified Hamiltonian. It can be shown that the pseudospins of the modified Hamiltonian are of the form, $\psi = \Big(\psi_1\,\,\,\psi_2\Big) = \Big(1 \,\,{\mathrm{exp}^{i\theta}}/{f(\beta,k_\mathrm{F})}\Big)$ where $f(\beta,k_\mathrm{F}) = \beta k_\mathrm{F}+\sqrt{1+\beta^2k_\mathrm{F}^2}$, independent of the angle ($\theta = \mathrm{tan}^{-1}{k_y}/{k_x}$) and therefore does not distort the chiral properties.

Transverse magnetic field (TMF)s have been used in the past to study various transport phenomena and characterize surfaces and interfaces \cite{tsoi1999}. In a recent experiment\cite{Herrero_focusing_NP13}, a TMF is used to focus electrons in a monolayer graphene device. The device geometry and biasing scheme used in our simulation are shown in Fig. \ref{fig:focusing}(a). Electrons are injected from contact with a current source 
$a$ and collected at contact $b$. In presence of 
a magnetic field $B$, an electron follows a circular path inside the graphene 
channel with cyclotron radius $r_c$ and is directly focused from 
contact $a$ to contact $c$ when $2r_c = L$, where $L$ is the distance between contacts $a$ 
and $c$. This is the resonance condition where the voltmeter registers a 
large voltage. Thus, the magnetic field required for the focusing is
\begin{eqnarray}
    B = m\frac{2\hbar\sqrt{\pi n}}{qL}
    \label{eq:B}
\end{eqnarray}
where $n$ is the density of electrons in the graphene channel and $m$ is 
an integer. For $m>1$ the electron reaches contact $c$ after skipping along the edges through multiple specular reflections.

NEGF simulations for the multi-terminal device are shown in Fig. \ref{fig:focusing}. The 
contacts are modeled using the self-energies of semi-infinite graphene 
ribbons. The current at contact $\alpha$ is calculated using the multi-terminal
Landauer-B\"{u}ttiker formalism \cite{buttiker1988,datta_97},
\begin{eqnarray}
    I_\alpha = \frac{q}{h}\sum_{i\neq\alpha}\int dE T_{\alpha i}(E)\left[f(\mu_{\alpha}) - f(\mu_i)\right]
    \label{eq:Ialpha}
\end{eqnarray}
where $f$ is the Fermi function, $\mu$ is the electro-chemical potential of 
the contact and $T_{\alpha\beta}$ is the total transmission between contact
$\alpha$ and contact $\beta$. The transmission is calculated using the Fisher-Lee formula \cite{datta_97},
\begin{eqnarray}
    T_{\alpha\beta}(E) = {\bf Tr}\{\Gamma_\alpha G_{\alpha\beta}^R\Gamma_\beta G_{\beta\alpha}^R\}
\end{eqnarray}
where $G^R$ is the retarded Green's function and 
$\Gamma_{\alpha,\beta} = i(\Sigma_{\alpha,\beta}-\Sigma_{\alpha,\beta}^\dagger)$ 
is the broadening from contacts ($\alpha,\beta$) with $\Sigma_{\alpha,\beta}$ being the corresponding energy dependent
self-energy matrices.
For computational efficiency, the retarded Green's function $G^R$ was 
obtained using the recursive Green's function algorithm\cite{alam2005} and the 
self-energies were calculated using the decimation method\cite{galperin_02}.
The real space Hamiltonian matrix for the channel and the contacts were
obtained using Eq. (\ref{eq:H}) and the effect of magnetic field was
included using the Peierls substitution\cite{sajjad_13}.
To calculate the voltage at contact $c$, we set $I_a$ = $-I_b = I$,
$I_c = 0$ and $\mu_a = E_\mathrm{F}$ then solve Eq. (\ref{eq:Ialpha}) for $\mu_c$
and $\mu_b$ where $E_\mathrm{F}$ is the Fermi level obtained from the electron 
density $n$. Then, $V = \mu_c - \mu_b$ and resistance $R\equiv V/I$. 
The whole procedure was repeated for each $n$ and $B$.

Fig. \ref{fig:focusing}(c) shows the Resistance $R$ as a function of the 
electron density $n$ and magnetic field $B$. The color map was generated 
using the quantum mechanical (NEGF) approach and the the dashed lines 
were computed using the semi-classical formula Eq. \ref{eq:B}. The dashed 
lines and the bright bands represent the focusing of electrons to contact $c$.
These results agree well with the experiment\cite{Herrero_focusing_NP13}.

As a final example, we show how our model can interpolate between the ballistic and diffusive limits. We model the transport with impurity scattering by using a sequence of Gaussian potential profiles for the scattering centers \cite{klos_10}, $U(r) = \sum_{n=1}^{N_\mathrm{imp}} U_n
\exp{(-{|r-r_n|^2}/{2\zeta^2})}$ that specifies the strength of the impurity 
potential at atomic site $r$. ${r_n}$ are the positions of the 
impurity atoms and $\zeta$ is the screening length ($\approx$ 3nm for long 
range scatterers). The amplitudes $U_n$ are random numbers following a 
Gaussian profile \cite{sui2011}, $N_\mathrm{imp}$ is the impurity concentration. 
With $U$ added to $H$ (potential landscape shown in Fig. \ref{diff}a-b), we 
study the evolution of electron transport in graphene from ballistic to diffusive (for varying $N_\mathrm{imp}$. 
Fig. \ref{diff}c shows the graphene 
conductivity at various channel carrier densities for several impurity 
concentrations. This time we keep the two ends of the graphene device at constant doping to capture the contact induced doping in graphene. This produces electron-hole asymmetry in the ballistic limit due to formation of $pn$ junction near the contact. At high impurity concentration, the contact resistance becomes less dominant compared to the device resistance and the 
electron-hole asymmetry washes out, similar to what is seen in experiments \cite{hollander2013}. 
Furthermore at the diffusive limit, $\sigma$ becomes proportional to $n$ for a sample dominated by long 
range scatterers and can be fitted with a carrier density ($n$) independent mobility. As we lower the impurity concentration towards the ballistic limit, the graphene conductance ($G$) becomes proportional to the effective doping $E_\mathrm{F}$, $G = {4q^2WE_\mathrm{F}}\mathcal{T}/{(\pi\hbar v_\mathrm{F})}$ leading to a sub-linear $\sigma \propto \sqrt{n}$), where the transmission $\mathcal{T}$ is determined by the metal-graphene contact. Such an evolution of electron  transport in graphene has been verified in experiments \cite{chen_08}. 


In conclusion, we have shown that for massless Dirac fermions, an additional 
quadratic term in the Dirac Hamiltonian not only circumvents the Fermion 
doubling problem in a spatial lattice but also has a huge computational advantage
over the atomistic tight binding model. In particular, we have shown that 
the modified Hamiltonian results in an extremely small matrix on a real space 
square lattice. As a result, the Hamiltonian is orders of magnitude faster 
than the tight binding Hamiltonian when used in band structure and quantum 
transport simulations. We applied this Hamiltonian for micron scaled graphene
devices to study magneto-transport and electron transport in ballistic and in diffusive limit. Although only graphene is considered here, it is applicable to any other Dirac materials like 
topological insulators and can be used to calculate the spin current \cite{habib2015} as well. 

\acknowledgements{This work was supported by the financial grant from the
NRI-INDEX center.}
\bibliographystyle{apsrev4-1}
\bibliography{sajjad_jab,BIBLIOGRAPHY}

\end{document}